# Generative AI and Its Educational Implications

Kacper Łodzikowski,

Adam Mickiewicz University, Poznań, Poland

klodziko@amu.edu.pl

Peter W. Foltz

University of Colorado

Peter.Foltz@Colorado.edu

John T. Behrens

University of Notre Dame

jbehrens@nd.edu

**Introduction**

Over the last 50 years, each decade can be considered to roughly align with the application of transformational technology that dramatically impacted daily life across societies. The 1970s brought the semiconductor, and the 80s the personal computer. The 1990s brought the World Wide Web, the commercial use of the Internet, open source software sharing models, and open standards for technology diffusion. The 2000s brought online search, e-commerce, and the scaling of the ideas and tools from the previous decade, fuelling the evolution of the passive consumption-oriented Web 1.0 into the increasingly interactive and participatory Web 2.0. The 2010s witnessed the proliferation of mobile devices and the growth of social media platforms, both of which contributed to the unprecedented accumulation of digital data. That decade also brought dramatic advances in data science and computing concerning the application of artificial intelligence (AI) for prediction and classification. This has led to a broad range of commercial applications of AI, such as virtual assistants enhanced by natural language processing (e.g., Amazon's Alexa), nearly autonomously driving cars, as well as decision support systems for healthcare, finance, and many other industries.

Writing this chapter in the spring of 2023, we believe we are on the cusp of the next wave of transformational technology that, like previous waves, appears both fantastical and naturally progressing. We have little doubt that when the 2020s are characterized in the future, they will be described as the 'decade of AI'. While AI has been evolving for decades, the field has taken a significant leap forward in the last year with the general availability of generative AI. This new area of AI is a collection of technologies in which computer systems use AI techniques and large amounts of data to generate texts, images, sounds, videos or their combination. The most well known product from this class of technologies is ChatGPT, which became a near overnight sensation upon its release on the last day of November 2022 (Brockman 2022). It reached 100 million monthly active users in just 2 months (Hu 2023). Nearly simultaneously, its underlying functionality was implemented in a wide range of technology products produced by Microsoft with similar functionality soon followed by Google and others. While less widely covered, technical capabilities for image generation from text instructions have also made dramatic improvements. Tools such as Midjourney allow users to generate photorealistic images that are sometimes impossible to distinguish from actual photographs, creating economic opportunities and social challenges.

This current wave of generative AI tools is different from prior waves in a number of practical ways. First, because it is building on top of prior technological waves, the rate of functional improvement is dramatically faster than we have seen in the past. The number of new articles referring to ChatGPT in the open-access paper repository arXiv grew from 25 in January 2023 to 772 in June 2023. This growth is enabled by the ability of researchers to 1) easily access state-of-the-art AI tools, 2) communicate with others widely and rapidly regarding findings, and 3) benefit from prior advances in open-source software characteristics including common libraries, languages, and systems for sharing (e.g., the computer code repository GitHub). While we believe the impact of generative AI as an underlying platform for many activities will match that of the Internet in the next decade, the rollout and impact will be dramatically faster as the AI evolution benefits from the existence of the Internet and related technologies, which the Internet itself could not benefit from.

Second, this wave supports specialized behavior of text generation (i.e. writing) and image generation (i.e. visual communication) in ways previously only attributable to humans and useful in the daily life of a great swath of society. Historically, automation technologies have largely affected blue-collar workers, especially in such industries as agriculture, manufacturing, and administration—and this was also the case with computerisation in the second half of the 20th century (Frey and Osborne 2017). And while the past two decades of AI advancements have not negatively impacted the overall job market due to AI adding some jobs on top of replacing others (Handel 2022), generative AI is different in that it is also expected to also impact white-collar jobs by automating such activities as sales and marketing content creation, customer service, or software development (McKinsey 2023).

Third, this widespread impact brings numerous social conflicts and confusions between the activities of humans and machines that we have not seen before. For example, while autonomous vehicles have been available for years, they are experienced by few and their impact is not perceived widely, not least due to their cost or regulation. However, the ability for virtually everyone to use freely-available, unregulated AI tools to generate writing or images whose provenance may be indistinguishable from human artifacts upends numerous social expectations and norms. This has already led to discussion about academic integrity (Cotton et al. 2023) and AI literacy (Anders 2023), confusion over the use of some tools in highly sensitive legal situations (Weiser 2023), and even concerns about the long term impact on modern societies writ large (Lukpat 2023).

In light of these rapid and significant changes, this chapter aims to provide the reader with an overview of text related generative AI technologies in four sections. First, we provide an overview of how AI relevant to education has evolved and a gentle introduction to how current technologies work. In the second section we discuss how such systems can be, and are being, applied in learning contexts followed by a third section in which we note a number of the larger societal issues that will impact AI in education. We conclude the chapter with recommendations for educational researchers. While generative AI is a class of software that includes a broad range of systems including text-to-text generation, text-to-image generation, text-to-video and other combinations, in this paper we focus on text-text generation (such as ChatGPT) as that is the area with which most readers will be familiar and which, we anticipate, will have the broadest base of use in education in the near future.

**Understanding language-based generative AI systems**

To help the reader appreciate the complexities of generative AI systems, we start with an overview of how the design of AI systems has evolved over the years. Since its birth after World War II, the field of AI has seen several cycles of growth and stagnation, often described in terms of technological breakthroughs and funding for commercial and academic endeavors (see e.g., Russell & Norvig 2022).

*Previous-generation AI: From rule-based to data-driven*

The first wave of AI systems, between the 1950s and the late 1980s, relied on domain experts to manually encode knowledge into a set of rules for logic-based decision-making. An early example of this approach is MYCIN (Shortliffe 1976), a dialogue-based system designed to aid physicians in diagnosing and treating bacterial infections. By posing close-ended questions to the physician, MYCIN simulated human expert decision-making using explicit rules (e.g., "If the patient is febrile, apply drug X") and experience-based heuristics (e.g., "If the blood test shows X and Y, then it is moderately suggestive that the bacteria is Z"). Although such expert systems performed satisfactorily in well-defined domains and straightforward cases, they struggled with complex real-world scenarios that fell outside their predefined rules and included elements of uncertainty. Moreover, the labor-intensive process of encoding the vast knowledge of human experts into hundreds of rules was a bottleneck. In the realm of education, this wave introduced intelligent tutoring systems that could dynamically track student knowledge, apply contextual tutoring strategies, and provide scaffolded support (e.g., Anderson et al. 1995; Sleeman & Brown 1982). For example, a variant of MYCIN called GUIDON (Clancey 1984) could engage a student in a mock dialogue about a patient's condition and give feedback on the chosen treatment. The system was proficient in remembering and applying rules and could understand and analyze the learner's input within the boundaries of its hand-programmed knowledge.

However, it struggled with evaluating ambiguous cases and was unable to create novel solutions not encoded in its database.

In the late 1980s, the second wave of AI introduced a shift from rule-based systems to data-driven machine learning systems. In this approach, the system identifies meaningful patterns in historical data and uses those patterns to generate rules for automated decision-making in the future. The role of the domain expert shifts from specifying the logical beliefs of experts to collecting relevant real-life data and pre-selecting (or 'engineering') data features (also called 'variables' or 'attributes' in the social sciences) likely to predict the outcome in question. Then, an algorithm ingests the features and goes through a cycle of 'learning' (also known as 'training') to produce a model which is used for prediction (also known as 'inference'). In the education space, this wave brought commercial-grade systems for automated formative and summative assessment. For example, the Intelligent Essay Assessor (Foltz et al. 1999) removed the burden of manual essay grading from teachers and provided learners with just-in-time feedback. Compared to a first-wave AI system, such a system was more robust in analyzing and evaluating learners' work products. This is because it no longer relied on manually crafted expert rules that could not cover the wealth of real-life situations.

The third wave of AI (from around 2011) introduced deep learning as a subfield of machine learning. Deep learning systems learn from data without requiring explicit feature engineering by domain experts. For example, to recognize handwriting, a deep learning system only needs a sufficiently large number of labeled pictures representing each handwritten character. This is made possible by a family of algorithms known as artificial neural networks, which are roughly inspired by the interconnected neurons in the human brain. In the 2010s, the capabilities of these deep learning models catalyzed the development of consumer-grade AI systems, such as automated labeling of photos on social networking platforms, speech recognition on mobile devices, and automated translation across multiple languages. As these AI-enabled tools became widely available, they propelled AI into the public's consciousness. Within the education industry, a new breed of self-study AI companions incorporated deep learning to offer more natural ways for the human to interact with the machine. For example, using Aida Calculus (Pearson Education 2019), learners could take pictures of their handwritten math problems to get step-by-step feedback on the process rather than just the final answer.

*Educational applications of previous-generation AI*

At this point, we can see that attempts at applying AI in education are not new. Previous-generation AI-based educational systems have been deployed for large-scale assessment, classroom and individual tutoring systems, and teacher support. Studies have shown that intelligent tutoring systems can raise student test scores 0.66 standard deviations over conventional classroom training and be as effective as expert tutors (e.g., D'Mello & Graesser 2023; Kulik & Fletcher 2016; VanLehn 2011). AI-based automated essay scoring has been used operationally since the early 2000s to grade high-stakes exams as well as provide students with instant feedback (e.g., Yan et al. 2020).

While successful, however, these AI-based educational approaches have had strong limitations. First, they have often been rule-based or trained on specific topics which can be inflexible and difficult to adapt to new situations. Second, they had limited natural language processing capabilities, resulting in more stilted, non-human-like language interactions. This has also limited systems to focusing less on higher-order thinking skills such as reasoning, argumentation, or collaboration. Third, while there has been research on multimodal processing, few incorporated modalities such as eye-gaze, gestures, facial reactions, or emotion detection, which can provide a deeper understanding of the learning context. Finally, automated educational systems have been expensive to build, often requiring collecting hundreds or thousands of hours of student interaction data to train models for specific domains or have needed content experts to code knowledge and design the interactions. Thus, while much has already been achieved in AI-driven educational systems, it is important to consider what generative AI enables that can accelerate the advancement.

*Pathway to generative AI*

The current wave of generative AI (originating around 2017) can be considered an extension of deep learning. For the past two decades, the AI research community has demonstrated that the quality of machine learning models, including deep learning models, tends to improve with their 'scale', defined as the amount of data available for model training and the computational resources required to process that data (e.g., Halevy et al. 2009, Sun et al. 2017). However, researchers soon encountered a time, cost, and quality bottleneck in the form of data collection and labeling. For example, building a neural network for detecting toxic social network posts required a dedicated data team to meticulously label thousands of historical posts as toxic or non-toxic, so that the AI could mine those examples for patterns and create generalized rules for classifying future posts. This forced a shift in approach: instead of relying on painstakingly curated datasets, researchers began experimenting with large, unstructured, and unlabeled datasets. They quickly turned to Internet-derived text, such as web pages, online encyclopedias, discussion forums, or digital books. This coincided with the development of a new neural network architecture called the transformer (Vaswani et al. 2017) which allowed for faster processing of large text files. Coupled together, these two developments paved the way for 'large' language models we see today.

At the training stage, the goal of a language model is to find patterns in texts in order to learn language patterns. In the 2000s, a language model could predict and correct words typed in a text message. In the 2010s, a deep learning-based language model could use its capability to predict language to generate grammatically-correct passages in the style of Shakespeare plays (Karpathy 2015)—albeit coherence degraded with longer generated texts. A breakthrough came with Generative Pre-trained Transformer-2, or GPT-2 (Radford et al. 2019), which was able to not only generate plausible language but also perform a wide range of tasks, such as document summarization, question answering, or translation. We should note that, even though GPT-2 could carry out a surprisingly wide array of tasks, its performance did not exceed that of humans or specialized AI systems of the time. For example, its cross-language translations were of lower quality than those of a specialized AI-based translation system developed for years using dedicated translation datasets. However, it demonstrated the feasibility of moving away from the established practice of developing multiple narrow-domain language processing systems towards a new paradigm of developing a single general-purpose system, or a 'foundation model'. GPT-2 was followed by GPT-3 (Brown et al. 2020), which approximated human-level performance on certain tasks and which was used, with modifications, in ChatGPT (OpenAI 2022). In the spring of 2023, ChatGPT was upgraded with GPT-4 (OpenAI 2023), which exceeded human-level performance on certain tasks (Bubeck et al. 2023), including beating specialized translation systems (Jiao et al. 2023).

*Capabilities of current-generation large language models*

While the capabilities of present-day large language models are still being explored, there are two main characteristics that distinguish them from previous-generation language models. The first is that they learn tasks from training data without supervision, that is, without humans specifying the things the models should learn apart from their basic goal to learn to generate sentences word-by-word. This unsupervised task learning capability emerged because the Internet is a treasure trove of real-life task demonstrations, and feeding a sufficiently large amount of such data into a transformer model allows it to learn not only the structure of language(s), but also the characteristics of the featured tasks. For example, if the training dataset includes text sources that feature the same sentences in English and French (e.g., language learning textbooks, fan translation websites, multilingual versions of the same document), then the language model will not only learn how to generate plausible English and French, but also what humans mean when they ask for a 'translation'. The unsupervised nature of training means that these models have learned to pick up more nuanced features of human language uses and contexts manifested in the data, such as sarcasm, sentiment, or cultural references. Consequently, current-generation language models not only complete tasks, but do so in a more human-like and contextually-appropriate way than ever before.

The second foundational capability is a human-like ability to learn how to perform tasks according to text-based commands, or 'prompts'. For example, while training an earlier-generation AI model to identify toxic social media posts would have required showing it thousands of examples, one now can provide a large language model with just a few example posts (real or fabricated) and it should be able to classify future posts based on those examples. This capability emerged because large language models were trained on increasingly larger and more varied text datasets, which allowed them to create representations of human reasoning and behavior demonstrated in the data, including learning from instructions. A trained model holds information that allows the production of language writ large, along with information on how to carry out tasks. When this general capability is combined with new, more specific information, such as examples of toxic posts, the system behaves in a manner that looks as if the system has integrated the principles of broad reasoning with specific information. The user can provide supporting instructions on how to execute the task and provide the model with feedback on how it should adjust its outputs. This practice of 'prompt engineering' enables people without extensive programming skills or computational resources to effectively 'program' their own copy of an AI model. In a way, this brings us back to the era of humans encoding their knowledge and preferences into AI systems, albeit not through manual programming of handcrafted rules, but by providing instructions and examples in natural, everyday language.

*From research to application*

As the general-purpose capabilities of large language models grew, AI developers began to adapt them to more specific uses, such as chatbots. For example, ChatGPT (OpenAI 2022) does not use the base GPT-3 model alone, but combines it with additional software and modified model layers focused on conversational interactions. The modification was needed because GPT-3, while capable, is not fully aligned to societal expectations due to its propensity for bias, toxicity, and misinformation (e.g. Lucy & Bamman 2021; Weidinger et al. 2022). This is because the model does not actually understand the text it generates in the same way that humans do—it mainly mimics previously seen texts. The performance of the model varies significantly according to the type of task given and its relationship to the training data. When it receives a prompt about a topic that was only briefly mentioned in its training data, it may generate text that is not aligned with reality. In contexts where creativity or novelty are valued, this may be a valuable characteristic. In contexts where facts are involved, however, such errors are called 'hallucinations' and are typically disregarded—though it is incumbent upon the human end-user to make the distinction. Accordingly, in many contexts the systems require targeted fine tuning to teach the system the most important information until it reaches the necessary threshold of accuracy or sustained human (or other computing agents) in the loop for risk mitigation.

To align models with human expectations of task performance, AI developers employ a few key techniques at the intersection of computer science and data science. One such technique is instruction tuning, which involves training the model on smaller curated dataset that consist of prompts and corresponding desired outputs, such as examples of how to give helpful and safe relationship advice (e.g., Zhou et al. 2023). Another technique is reinforcement learning from human feedback (Ouyang et al. 2022). In this approach, the model is trained to adapt its responses based on feedback it receives over time. For example, if the model provides a response that is factually incorrect or inappropriate, it can be corrected, and that correction is factored into its future interactions. A more fundamental approach to minimizing harmful model behavior is to train it on a dataset of higher quality, such as websites known for factual accuracy (e.g., Touvron et al. 2023).

Other strands of research focus on finding strategies for interacting with models to reliably obtain truthful outputs (e.g., Bubeck et al. 2023). For example, when consulting a generative chatbot for answering fact-based multiple-choice questions, it may be more beneficial to ask the model to explain its reasoning before trusting its correct answer choice (Bowman et al. 2022). Another emerging practice involves prompt engineering templates, that is, proven strategies that maximize the chance of obtaining desired outputs, such as asking the model to generate probing questions until it collects adequate information to deliver a relevant response (White et al. 2023). The field is also witnessing developments that enhance the capability of generative AI systems to retrieve information from trusted sources. For example, systems such as Toolformer (Schick et al. 2023) can call upon other models or databases to provide factual information or perform complex calculations.

As we look forward, the horizon of large language model research is widening beyond just natural language. Models trained on computer code datasets, such as Codex (Chen et al. 2021), can be prompted to generate entire computer programs. Multimodal models, such as Kosmos-1 (Huang et al. 2023) or GPT-4 (OpenAI 2023), can process and generate more than one type of data. For example, the user can upload an image and the model can describe it. This broadens the ways in which they can understand and interact with the world, addressing one of the fundamental concerns behind text-only language models, namely that their understanding of reality is only grounded in what can be represented in text form.

**Opportunities and applications**

*Interaction and assessment in education*

To examine where advancements can be made from generative AI, we consider two key facets that comprise education and work together to create an effective educational experience: *interaction* and *assessment*. Much of education comprises a multiway multi-modal *interaction* between learners and agents (e.g., other learners, instructors, or responsive educational artifacts such as ITSs). An agent can be conceived as a 'system situated within and a part of an environment that senses that environment and acts on it, over time, in pursuit of its own agenda and so as to effect what it senses in the future' (Franklin & Graesser 1997). For example, in a learner-instructor face-to-face dialogue, an instructor can question the learner, dynamically adapt their responses to the level of the learner, and provide visual, auditory, gestural, and emotive responses. Thus, the agent can sense the state of a learner and choose responses that can be most effective for impacting a student's learning. On the other hand, much of education is an interaction between learners and static materials (e.g., books, manuals, web pages). For example, a book is carefully crafted by the author so that each paragraph follows coherently from the next with an organized structure that is oriented to providing new information at a rate that can be absorbed by the reader within their zone of proximal development (e.g., Vygotsky 1978). However, it does not adapt itself to differing learning contexts or learner levels.

In order to be effective in interacting with a learner, an agent must be able to perform an *assessment*. Assessment in education means being able to infer attributes of the learner through observation of their performances and activities in natural or controlled contexts (e.g., Behrens and DiCerbo 2014). It, therefore, provides the means for an agent to sense the environment (e.g., the state of the learner in relation to the learning situation) and guide how best to act upon the learner's state. Assessment is critical for evaluating learning, guiding instruction, knowing when to provide feedback, tracking progress, as well as measuring accountability of educational systems. Within digital environments, assessments can be embedded and integrated as part of the natural learning experience. Furthermore, assessments deriving from a variety of digital experiences can be combined to make inferences about student ability over longer time frames (e.g., DiCerbo & Behrens 2012). To accomplish this, the types of assessments must be aligned with the tasks being performed by the learners. While multiple choice and fill-in-the-blank type assessment items have been widely used and are easy to automatically score within digital learning environments, they often reduce the complexity of the activity to match the scoring format rather than considering how richer inferences can be extracted from more complex tasks and responses.

Analyzing information from complex performances, such as speaking, writing, and the logging of process data is difficult for both humans and computers. However, automating these assessments enables the integration of more complex performances within learning environments (e.g., Behrens et al. 2019). Over the past 30 years, there has been great advancement in applying AI for assessing writing, analyzing spontaneous speech within tutoring contexts, and mining process data (see Koedinger et al. 2015; Yan et al. 2020; Zechner & Evanini 2019 for reviews). These advances have allowed the development of more interactive learning systems in which the assessments are embedded as part of the performance. These systems include interactive dialogue-based tutoring, automated assessment of writing with instant formative feedback, and tracking and feedback on teams performing collaborative tasks. Yet, the AI-based assessment techniques that are used typically require collecting a large number of samples of student performance, hand-scoring them, and then using machine learning techniques to train an AI model to learn to score them automatically. This limitation has confined the applicability of AI to areas where data collection is straightforward, interactions can be hand-designed, and human coders can easily characterize performance.

The advent of generative AI, however, promises to greatly transform assessment methodologies, addressing many of the limitations currently faced in the field. Whereas automating assessment has required handcrafted models and training data, now, with its broad domain knowledge and ability to generate learning experiences through prompting, generative AI can be easily implemented by teachers and developers without advanced AI training and can be used in many domains. For one, when provided with spoken or written language from a learner as input, it can characterize multiple qualities of a learner's language and cognitive abilities. It can also integrate multimodal data, such as speaking, writing, facial emotions allowing for a more personalized understanding of a student's strengths and weaknesses. Additionally, generative AI can be instructed to adhere to a particular rubric, providing an objective and standardized means of evaluation through prompt engineering. By writing a carefully crafted prompt, an educational designer can instruct the AI to assess consistently, thereby reducing inconsistencies in grading that may arise due to limitations in human assessment capabilities, such as the need for training, and the requirement of continued human attention. Furthermore, assessment through generative AI not only provides a measurement but can also give meaningful explanations for each assessment, fostering understanding and transparency in the evaluation process. Lastly, different types of language models can be applied across different written and spoken languages as well as software code, making it a versatile tool in multilingual education environments as well as for learning programming skills.

*Generative AI for complex performances*

The advent of generative AI presents an opportunity to overcome the above-mentioned hurdles and provide agency for interactivity and assessment in educational technologies. The nature of the prior training of generative AI means that the automation of digital interactive learning experiences does not have to be as hand-crafted or developed through collection of large amounts of prior training data that is specific to the contexts. Thus, the foundation models in generative AI provide a means to jump off into new educational innovations, much in the same way that the Internet suddenly allowed data interchange, which resulted in many new forms of knowledge sharing which have become the primary means for communicating and collaborating. For example, prior generations of AI question/answering systems would have required painstaking training of the dialog system with numerous specific examples of acceptable or partially acceptable responses. The new large language models come pre-built with that language assessment functionality built in, thereby greatly accelerating the speed of development for many new systems.

Interactivity in digital environments will change through generative AI allowing the creation of more engaging, realistic learning experiences. First, as an agent, AI can take on roles that are much more like human-human interactions (see Office of Educational Technology 2023). While learners have formerly mostly written and made simple click responses with online systems, AI will allow in-depth interactive conversations through speaking and drawing, with the system able to respond conversationally with the dialogue adapted to the appropriate level of knowledge and language ability of the learner. Second, these agents further have the ability to assume different roles, such as a mentor, tutor, coach, peer teammate, as a student that needs teaching, or as an embedded simulation (e.g., Mollick & Mollick 2023). Each role may be optimized for different learning situations. For example, learners working with an AI teammate on a collaborative problem solving task can learn strategies such as how to construct shared knowledge and maintain team functions (see Graesser et al. 2018). By participating as a teammate, the AI-agent can both support the team of learners by serving as an example, but also monitor and adapt its responses to help improve the functioning of the team. Third, generative AI has the ability to generate information on the fly that is adapted to the learner's needs. Rather than choosing a static textbook that is written at the level of the learner, a learner can choose to read about a topic and interact with a system that generates content adapted to the learner. For a learner that needs to study mitosis in biology, an AI system can generate text explanations adapted to the student's background knowledge and reading level. Moreover, it can generate images, movies and simulations to further explicate the examples. It can further respond to various forms of communication including spoken language, written texts, and even facial emotions to continually adapt based on how well the learner is grasping the material.

Thus, generative AI opens the field of education to novel approaches to creating learning assessment contexts, evaluating the quality of responses and generating contextually appropriate feedback. We summarize this potential in Table 1, showing different types of multimodal language models, how they can provide interactive and/or assessment, the kinds of educational methods that can be applied, and potential educational applications that can result from them.

| Language model type | Interactivity and assessment | Sample methods | Sample application |
|---|---|---|---|
| Text-to-text | Create | Instructional material generated on the fly adjusted to learner level | Personalized textbooks |
| | | Generate contextual assessment activities e.g., multiple choice questions | Practice items with difficulty adjusted to student learning level |
| | Evaluate & feedback | Compare student written response to domain content and generate feedback | Writing practice for content areas with instant formative feedback |
| | Create & adapt | Act as roleplay participant, adapting character based on student prompts | Dynamic learning environments that facilitate integrated development of |

|  |  |  | critical analysis skills |
|---|---|---|---|
| Text-to-code | Create | Generate contextual assessment activities, e.g., logically-correct computer code with syntax errors | Software debugging practice |
| Code-to-text | Evaluate & feedback | Assess quality of student computer code and convert to description of errors | Instant assessment and deep conceptual feedback and training |
| Text-to-image | Create materials in new modality | Generate illustrations/diagrams based on textual descriptions | Visual aids for complex theoretical concepts |
| Image-to-text | Evaluate activity in new modality | Recognize handwritten math to provide step-by-step feedback | Pinpoint diagnosis of gaps for remediation |
| Text-to-speech | Create materials in new modality | Generate speech from automatically generated training materials | Interactive speech-based tutors for content domains |
| Speech-to-text | Evaluate activity in new modality | Interpret quality and accuracy of speech signal | Interactive dialogue for language learning and practice |

Table 1. A selection of multimodal language model types and their potential applications in education.

*Towards more engaging real-world learning experiences*

By combining interactivity and assessment, generative AI enables more engaging, natural learning experiences with a higher level of fidelity in measuring learner performance. This advancement not only facilitates a deeper understanding of students' abilities but also opens avenues for a broader range of real-world experiences through realistic simulations and embedded games. Moreover, the integration of generative AI holds great potential to increase the relevance of training in schools and workforce development programs. It can cater to a variety of higher-order thinking skills and domains, such as coding, critical thinking through writing, and teamwork, particularly in genres where learners have limited experience. For instance, generative AI can support learners in tasks such as crafting a compelling argument after reading multiple documents, effectively collaborating with team members, or simulating realistic interview scenarios. By immersing students in these practical performances, generative AI fosters skill acquisition and prepares learners for real-world challenges.

The approach also changes how we conceive of assessment and allows us to move towards a model of continuous assessment and learning. Instead of treating the educational process as a set of separated learning experiences and summative assessments, all assessments are embedded in the learning activities with real-time feedback. These kinds of instantaneous feedback loops using AI have proved advantageous for learning higher-order skills. For instance, in learning to write in content domains, AI-based feedback on learners' content knowledge and writing skills allows learners to iterate with the computer to refine their essays before submitting them to teachers. This has resulted in faster learning of the domain knowledge and writing skills as well as providing a function for assessing thousands of drafts and alerting the teacher to students who are struggling (e.g., Foltz et al. 2013). Thus AI-based tools open new models of education for both students and teachers. For students, they can iterate with agents, practicing and learning. Teachers can rely more on formative assessment loops where they are still the guides of the learning process, directing when and how the AI will be used, but able to be continually informed about the state of student learning, and able to intervene and engage with students. As such, the goal of AI in the classroom is not to replace the teacher, but to empower them with tools that increase their effectiveness.

**Challenges to implementing, deploying, and using generative AI-based educational tools**

*Choosing the right tool for the job*

The application of large language models in education is a nascent field, and many aspects of their behavior have yet to be sufficiently explored. The availability of large language models such as the GPT family and the ease of prompt engineering allow people to rapidly develop systems for assessment and learning, such as chatbots. However, just because such a system is built on a generative AI model that has proven effective for some online tasks does not mean that the approach will be effective as an educational tool. As more and more companies release new models, there arises a concern about determining which one is the most effective for a particular educational experience. This is because the models differ in fundamental assumptions, such as the quantity and quality of training data or any additional alignment of the model's capability to human expectations in a given context and domain. Another set of considerations involves the learning context, the specific needs and preferences of the students, and the objectives of the course or program. Therefore, there are a number of challenges that still need to be addressed for implementing, deploying, and using generative AI-based educational tools at scale.

*Data bias and design transparency*

One of the principal concerns of using AI-based models is that they reflect the data they are trained upon. The quality of the data that is used in training is crucial. As machine learning models learn from data, any inaccuracies, omissions, or biases within the data can be reproduced and magnified in the AI's behavior. Bias in AI can take many forms and can result in unfair or inequitable outcomes (e.g., Baker & Hawn 2022). For example, if algorithms are not trained on a diverse set of student responses, they may disproportionately penalize or reward certain ways of communicating, thinking, or problem-solving (e.g., Kwako et al. 2023). This can have different kinds of effects on students from various cultural, linguistic, or socio-economic backgrounds.

However, ensuring the accuracy, diversity, and breadth of data is a significant challenge. As of today, educational technology developers do not have control over how large language models are trained or what kind of data was used in the training. Moreover, most model providers are not transparent regarding the design of their systems and do not provide guarantees against bias. This is a major concern in educational contexts, which require fair and equal opportunities for all students. Addressing this challenge requires continuous efforts in bias detection and mitigation in both the data and algorithms used in AI systems. Indeed, developers of educational systems may still need to test and certify their systems across wide ranges of inputs to assure that biases are mitigated, or at least are known so that the system is only used in contexts for which the models are appropriate.

*Algorithmic explainability and propensity for misinformation*

Assuring the quality of generative AI systems and their outputs is another significant challenge. As AI often functions as a 'black box', it is difficult to understand precisely why a model is making a certain decision or prediction. This is particularly problematic when we consider the psychometric properties of an assessment, especially validity and reliability. Traditional methods of evaluating those characteristics may not directly apply to generative AI systems. Most likely, new methods will have to be developed that consider changes in the assessments and the nature of the learning environments (e.g., von Davier et al. 2021).

The issue of model quality is related to another key challenge, namely the potential for large language models to 'hallucinate', that is, generate information that seems plausible but is incorrect or misleading. This can be especially harmful in educational contexts, where accuracy of information is paramount. And while hallucination seems to be caused primarily by the word-by-word nature of text generated by transformer models, addressing it systemically requires not only better models but also robust systems for verifying and validating AI outputs.

*Introducing and maintaining standards for generative AI in education*

These challenges illustrate that generative AI-based systems cannot be deployed on learners without a significant amount of research, testing, validation, and human oversight. Indeed, the field will need to internally police itself with standards that espouse transparency and explainability around the methods. This includes being open about how the models were developed, tested and validated, and providing information on their intended use and limitations in their educational context (e.g., Mitchell et al. 2019). Concurrently, the field will need to continually incorporate external guidance to help steer ethics in this field, such as the European Union's ethical guidelines on the use of AI and data in teaching and learning and education (European Commission 2022). In certain areas of the world, such guidelines will be reinforced by formal transparency regulations (e.g., European Parliament 2023).

While AI may provide greater autonomy for learners and instructors, it need not take human instructors out of the educational process. It may change how they interact with learners and computational systems. It may change how they select educational material or structure their courses. It may change the kinds of information they receive about learners and allow them to focus more on those learners in need. But humans will still play a critical role in orchestrating how the AI is applied to best impact learners. Indeed, we see that our future world will require a human-AI partnership in which each provides their specialized capabilities and collaborate, resulting in something more educationally effective than either working individually (e.g., Hellman et al. 2019). Thus, we advocate for a human-in-the-loop approach throughout the development and use of AI-based systems in education, as it is the human teacher who will act as the ultimate regulator.

**Challenges for educational ecosystems**

In the previous sections, we have primarily focused on technological advances and their impacts on providing educational support through technology. In this section, we aim to highlight a few areas in which education itself will experience second-order effects stemming from the larger societal changes that these technologies will bring, independent of their educational applications. We will briefly discuss this in relation to the evolution of the workforce and daily life activities, the transformation of modes of communication and its impact on issues of trust, and finally, the evolution of social norms. These represent significant societal changes that will undoubtedly have a profound impact on education on a grand scale.

*New work, new curricula*

Perhaps the clearest signal from the new AI is that digital technologies can now perform many tasks that only humans could do just a very short time ago. For example, systems such as ChatGPT can write quality computer code at a scale that is already changing the landscape and best practices of software development. When large language models are fine-tuned for specific tasks, such as writing computer code or answering legal questions, their outputs are often sufficient as first-draft work products to be inserted into production workflows. While the prior phases of AI and related technologies such as robotics replaced mainly certain physical tasks, such as elements of the automobile assembly process, the current wave is squarely focused on language-based, and therefore cognitive, activity. In fact, higher-order thinking skills required for such jobs as accounting may be more impacted by AI in the foreseeable future than sensorimotor skills required for such jobs as housekeeping. This is backed up by historical data—despite their proliferation, autonomous robotic vacuum cleaners have not impacted housekeeping jobs (Handel 2022)—as well as the challenges of building AI-powered robots that will walk, sense, and act with the same dexterity as humans (Deranty & Corbin 2022). Even though predictions vary on how much cognitive work will be augmented by generative AI versus replaced by it, it is clear that the workforce of the future will have to master generative AI tools. They will also need to evolve their skills and foci, so that they complement rather than compete with the new capabilities of AI.

In the education realm, there will be many impacts at the administrative levels. First, curricula must change to prepare students for a rapidly evolving world. Otherwise, we risk preparing the next generation for a world that no longer exists. At present, many educational institutions are not well positioned to evolve their curriculum quickly, typically having well established curricula and faculty incentivized for long-term concerns. This leads to a second concern, that the manner in which faculty operate will need to change with regard to both the production of research and the conduct of instruction. In the same way that the shift from print to digital representation of knowledge via the Internet led to a sea change in the speed and quality of academic research, we expect to see a similar explosion of productivity. In instruction, not only must the curricula change, but also the modes of instruction must, and will change, as we discussed above. While tools to support this will unfold over the next few years, we are likely to see a period of disconnect where digitally-native students are out of sync with the understandings and practices of less technically-oriented faculty.

*New modes of communication and trust*

Aside from the pragmatic issues of workforce evolution and appropriate skill development, the availability of generative AI presents a fundamental challenge. Until now, while we have seen an ongoing increase in automation and the use of information technologies, it has always been, by and large, straightforward to distinguish between human-generated and machine-generated products. For instance, in the realm of images, we have a long history of understanding print and digital images as having a high verisimilitude to physical reality because photographs were designed for that purpose. However, current image generation technology, available to everyone through the Internet, can generate photo realistic images that are extremely difficult to distinguish from photographs. This suggests that the epistemic and social assumptions we bring to interpreting images need to be rethought. For example, while anticipating Donald Trump's indictment in 2023, a journalist used a widely-available AI tool to generate images imagining Trump's arrest (Belanger 2023). These images were widely reposted, often with viewers believing they were photographs, rather than computer-generated images. Similarly, in July of 2022, a generated picture of the Pentagon on fire was passed around the Internet with some attributing a drop in the stock market to the perceived 'news' (Polus 2023).

There is a fundamental issue that whereas earlier we could assume with high (though not perfect) confidence that images reflected physical realities, that assumption can no longer be held without question. We expect to see similar issues in text generation. Among the many interesting qualities of large language models are their ability to generate or regenerate text 'in the voice of…'. For example, ChatGPT produces the following opening sentences when asked to rewrite the previous paragraph in the voice of Thomas Jefferson:

'In the realm of workforce evolution and the cultivation of suitable skills, there arises a profound quandary concerning the advent of generative AI. Hitherto, while witnessing the continuous rise of automation and the integration of information technologies, we have generally been able to discriminate with ease between human-crafted endeavors and those wrought by machines.'

While this is an interesting linguistic and historical exercise that may have curricular implications for historical analysis, it could also lead to widespread fakery and political misinformation for historical figures as well as for current events. This is a watershed moment in how our societies will understand and react to attribution and provenance going forward, and how educational institutions will evolve to support them.

*The collaboration boundary and social norms*

The rapid introduction of generative AI into daily routines raises numerous questions about appropriate use which are not always straightforward to answer. As the fundamental differences blur between what humans and computers are capable of, fundamental questions of attribution and provenance are raised as well. At present, people are generally not required to cite the version of a grammar checker they use to manipulate the text of a passage for increased clarity. However, the additional cognitive-like functionality of large language models requires a more precise language about what 'what I have done' and 'what the computer has done', which is not evident at present. There is clarity on extremes, such as when an AI system writes an entire paper, or when it is used only as a research tool similar to searching the Internet. But in the middle, for example when the software has synthesized ideas or provided novel formulations, should the software be cited as a co-author or a tool in the same way statistical software may be cited in a quantitative analysis? We do not yet have answers to such questions but advise patience and generosity. There will be many perceived social transgressions and mistakes while social and professional societies evolve their understandings and practices.

**Conclusion**

We are at an inflection point in the relationship between computers and humans that has only been previously suggested in science fiction, with both utopian and dystopian implications. The behavioral capabilities of large language models and other forms of generative AI are evolving so rapidly that the technical leaders in the field frequently express surprise at these systems' behaviors; a fact that led some to sign a letter requesting a slow down in generative AI-related product development (Bengio 2023; Seetharaman 2023). Whether educational researchers are focused on the utopian or dystopian implications, building or using the technologies, or focusing on the social and ethical critiques, these technologies are impacting our societies and educational systems and must be actively engaged. Several recommendations follow.

First, educational researchers should start using the freely-available text-to-text generation tools as part of their ongoing personal or professional activities. Both the use of these tools and the tools themselves are rapidly evolving as products. For example, in March of 2023, ChatGPT was updated with the latest large language model from the GPT family, GPT-4, which allowed the general public to experience the improvements in model output quality. Almost immediately, another feature called plug-ins was introduced that allowed the system to connect to other software, such as Internet search engines, mathematical problem solvers, and travel databases, thereby uniting the 'large language brain' with access to real time data. These innovations will continue and these systems will evolve. The speed of technological evolution puts numerous social practices at risk and we encourage the scholarly community to engage with the technologies to help guide social evolution.

Second, educational researchers need to become conversant with the fundamental logics we introduce here. We can think of the role of the Internet in the evolution of organizations over the last 20 years. While not every organization became an 'Internet company', almost all organizations have become Internet-dependent. Similarly, while not all educational researchers need to become AI researchers, all researchers must know enough to evolve their research and teaching on the new AI platforms as appropriate. To help society in its social evolution with technology, and to take advantage of its benefits, we must achieve the required level of understanding and engagement.

Third, researchers must rethink their relationship to technology and its use. For many in education, technology is a niche topic for others to consider. We hope we have sufficiently communicated that the inflection point of technology infusion we are facing will change how society interacts with technology and how we as educators relate to technology. The concerns and opportunities we are facing involve curricular issues, psychological and social issues, computational and media issues and so forth. Technologies such as data science and machine learning are no longer topics but substrates to our daily lives and our educational research should reflect it.

Fourth, for those interested in computational aspects of education, this is both an exciting and challenging time. The speed of change in both research results and industrial applications is remarkable. The fact that large language models perform well at computer coding means that new support is available for those who want to enter the world of computer science. At the same time, for those with prior software experience, there are a flood of support tools for learning how to use and adapt open-source or proprietary AI software. A notable example is the model repository and cloud computing environment offered by Hugging Face (https://www.huggingface.co).

For both good and ill, the biggest limitation that educational researchers will face is their imagination. The new AI systems act, and are interacted with, in such novel ways that students and researchers with limited computational background may find it difficult to appreciate the opportunity to build something new. At the same time, those with engineering background may be limited by old conceptualisations of learning and assessment instead of reimagining what might be possible. Either way, we encourage all researchers to learn, experiment, and integrate their domain knowledge with these new developments.